# Giant Transverse Optical Forces in Nanoscale Slot Waveguides of Hyperbolic Metamaterials

Yingran He[1,2], Sailing He[2], Jie Gao[1] and Xiaodong Yang[1,*]

[1] *Department of Mechanical and Aerospace Engineering, Missouri University of Science and Technology, Rolla, MO 65409, USA*

[2] *Centre for Optical and Electromagnetic Research, Zhejiang Provincial Key Laboratory for Sensing Technologies, Zhejiang University, Hangzhou 310058, China*

*Corresponding author: [*]yangxia@mst.edu*

**Abstract**: Here we demonstrate that giant transverse optical forces can be generated in nanoscale slot waveguides of hyperbolic metamaterials, with more than two orders of magnitude stronger compared to the force created in conventional silicon slot waveguides, due to the nanoscale optical field enhancement and the extreme optical energy compression within the air slot region. Both numerical simulation and analytical treatment are carried out to study the dependence of the optical forces on the waveguide geometries and the metamaterial permittivity tensors, including the attractive optical forces for the symmetric modes and the repulsive optical forces for the anti-symmetric modes. The significantly enhanced transverse optical forces result from the strong optical mode coupling strength between two metamaterial waveguides, which can be explained with an explicit relation derived from the coupled mode theory. Moreover, the calculation on realistic metal-dielectric multilayer structures indicates that the predicted giant optical forces are achievable in experiments, which will open the door for various optomechanical applications in nanoscale, such as optical nanoelectromechanical systems, optical sensors and actuators.

Optical forces arising from the gradient of light field have been extensively employed to realize exciting applications for light-matter interactions, such as optical amplification and cooling of mechanical modes[1], actuation of nanophotonic structures[2-4], optomechanical wavelength and energy conversion[5], and optical trapping and transport of nanoparticles and biomolecules[6, 7]. It has been shown that optical forces can be remarkably enhanced with coupled high-$Q$ optical resonators where the circulating optical power is considerably amplified due to the long photon lifetime[8, 9]. Besides, such gradient optical forces can also be significantly enhanced through the compression of optical energy into deep subwavelength scale. Recently, strongly enhanced optical forces have been obtained in dielectric slot waveguides[10, 11] and hybrid plasmonic waveguides[12]. Since the optical field enhancement is proportional to the index-contrast at the slot interfaces, a material with a higher refractive index is desirable to further boost the optical force in slot waveguide structures. Metamaterials can be carefully designed to exhibit ultrahigh refractive indices[13-15], which are not available in naturally occurring materials at optical frequencies. Especially, hyperbolic metamaterials constructed with metal-dielectric multilayers supports huge wave vectors and therefore also ultrahigh refractive indices, due to the extreme anisotropy of permittivity tensor[16-19].

In this paper, we will demonstrate that the transverse optical forces in slot waveguides of hyperbolic metamaterials can be over two orders of magnitude stronger than that in conventional dielectric slot waveguides[11]. The mechanism of such optical force enhancement will be investigated both numerically using Maxwell's stress tensor integration, and analytically using a 2D approximation of the 3D slot waveguide system. Moreover, the relation between the optical force and the waveguide mode coupling strength is derived based on the coupled mode theory analysis[20]. The comprehensive understanding of the enhanced transverse optical forces in metamaterial slot waveguides will be very useful for nanoscale optomechanical applications, such as optical tweezers[21,

[22], optomechanical device actuation[23] and sensitive mechanical sensors[24].

Fig. 1(a) shows the schematic of the hyperbolic metamaterial slot waveguides. Two identical waveguides with width $L_x$ = 40 nm and height $L_y$ = 30 nm are separated with a nanoscale air gap $g$ along the $y$ direction. In each waveguide, the metamaterial is constructed with alternative thin layers of silver (Ag) and germanium (Ge). The multilayer metamaterial can be regarded as a homogeneous effective medium and the principle components of the permittivity tensor can be determined from the effective medium theory (EMT)[25]

$$\varepsilon_x = \varepsilon_z = f_m \varepsilon_m + (1-f_m)\varepsilon_d, \quad \varepsilon_y = \frac{\varepsilon_m \varepsilon_d}{f_m \varepsilon_d + (1-f_m)\varepsilon_m} \quad (1)$$

where $f_m$ is the volume filling ratio of silver, $\varepsilon_d$ and $\varepsilon_m$ are the permittivity of germanium and silver, respectively. $\varepsilon_d$ = 16, and $\varepsilon_m(\omega) = \varepsilon_\infty - \omega_p^2/(\omega^2+i\omega\gamma)$ from the Drude model, with a background dielectric constant $\varepsilon_\infty$ = 5, plasma frequency $\omega_p$ = 1.38×10$^{16}$ rad/s and collision frequency $\gamma$ = 5.07×10$^{13}$ rad/s. Fig. 1(b) shows the dependence of the permittivity tensor on the silver filling ratio $f_m$ at the telecom wavelength $\lambda_0$ = 1.55 μm. All the components of the permittivity tensor will grow in magnitude as the filling ratio increases. For instance, the permittivity tensors of hyperbolic metamaterial are $\varepsilon_y$ = 29.2 + 0.12i, $\varepsilon_x = \varepsilon_z$ = -39.8 + 2.1i for $f_m$ = 0.4, and $\varepsilon_y$ = 76.3 + 1.4i, $\varepsilon_x = \varepsilon_z$ = -81.7 + 3.6i for $f_m$ = 0.7, respectively. It should be noted that the wavelength $\lambda_0$ = 1.55 μm is merely used as an example throughout the paper, and this design can actually work in a broadband frequency range due to the non-resonant nature of hyperbolic metamaterials[26].

The strong mode coupling between the two closely spaced waveguides will generate mode splitting of the individual waveguide mode and result in two eigenmodes; one is the symmetric mode (denoted by $M_s$) and the other is the anti-symmetric mode (denoted by $M_a$). The effective indices $n_{eff,z} \equiv k_z/k_0$ and the propagation length $L_m \equiv 1/2\text{Im}(k_z)$ corresponding to the two eigenmodes are obtained by finite-element method (FEM) with the software package COMSOL (where $k_0$ is the wave vector in free space and $k_z$ is the wave vector along the propagation direction $z$). Fig. 1(c) and (d) show that the dependences of $n_{\text{eff},z}$ and $L_m$ on the gap sizes are distinct for the two eigenmodes. As the gap size $g$ shrinks, $n_{\text{eff},z}$ grows dramatically for mode $M_s$ but decreases slightly for mode $M_a$. In fact, the magnitude of the effective index variation is equivalent to the mode coupling strength between two identical waveguides[12], and therefore the distinguished variations of effective indices for two eigenmodes imply a strong coupling strength for mode $M_s$ and a weak coupling strength for mode $M_a$. Furthermore, the opposite effective index variations for two eigenmodes indicate an attractive force for mode $M_s$ and a repulsive force for mode $M_a$. The propagation length $L_m$ decreases for mode $M_s$ and increases for mode $M_a$ as the gap size gets narrower, due to the tradeoff between the optical mode confinement and the propagation loss. The effective index and the propagation length for the unperturbed mode of the individual waveguide are shown in Fig. 1(c) and (d) for comparison (denoted by $M_0$).

The optical mode profiles of field components $E_y$, $H_x$ and $S_z$ for slot waveguides with $g = 10$ nm are shown in Fig. 2. As can be seen from Fig. 2(a), distinct behaviors in electric field $E_y$ are obtained for modes $M_s$ and $M_a$, where a strong (weak) electric field is localized in the gap region for $M_s$ ($M_a$) mode. It has been proposed that the optical field could be tightly confined and greatly enhanced in the nanoscale slot region due to the large discontinuity of normal electric fields ($E_y$ in our case) at the high-index-contrast interface[10]. Here we demonstrate that strong optical field confinement is achievable in the slot region for the symmetric modes, due to the constructively interfered electric field. While for the anti-symmetric modes, weak optical field confinement is obtained in the slot region. Fig. 2(b) shows that the magnetic fields are always tightly confined within the hyperbolic metamaterials for both eigenmodes due to the absence of magnetic response for the metamaterials at an optical frequency. Accordingly, a large amount of energy flow is guided in the slot region for the symmetric mode, while a negligible amount of energy is confined in the slot region for the anti-symmetric mode [see Fig. 2(c)].

The optical field confinement and optical energy squeezing within the slot region are directly related to the transverse optical forces exerted on each metamaterial waveguide, which can be evaluated by integrating the Maxwell's stress tensor $\bar{\bar{T}} = \varepsilon_0 \vec{E}\vec{E} + \mu_0 \vec{H}\vec{H} - \bar{\bar{I}}/2(\varepsilon_0|\vec{E}|^2 + \mu_0|\vec{H}|^2)$ around an arbitrary surface enclosing one waveguide[27], and the non-vanishing force component $f_{opt}$ along the $y$ direction is

$$f_{opt} = \frac{\oiint_S \bar{\bar{T}} \cdot d\vec{S}}{P_z} \cdot \vec{e}_y \qquad (2)$$

where $\vec{S}$ is a surface enclosing one metamaterial waveguide, $\vec{e}_y$ is the unit vector along the $y$ direction. Here the transverse optical force $f_{opt}$ is normalized to the total optical power $P_z$ confined in the coupled waveguides. The optical field intensity at the slot region directly indicates the magnitude of transverse optical forces between the two coupled waveguides, and therefore the optical force for the symmetric mode is expected to be much stronger than that for the anti-symmetric mode, due to the distinguished optical field distributions for the two modes (see Fig. 2).

Fig. 3 shows the effective indices $n_{eff,z}$ and the optical forces through the integration of Maxwell's stress tensor for both the symmetric mode and the anti-symmetric mode. $n_{eff,z}$ for the symmetric mode increases noticeably as the gap size shrinks [see Fig. 3(a)], implying a strong mode coupling strength between the two waveguides. Accordingly, the attractive optical force for the symmetric mode grows dramatically with the decreased gap sizes, resulting in optical forces up to 8 nNμm$^{-1}$mW$^{-1}$ for $L_y$ = 30 nm and 4 nNμm$^{-1}$mW$^{-1}$ for $L_y$ =80 nm [see Fig. 3(b)], over two orders of magnitude larger than that in a dielectric slot waveguide[11]. On the contrary, $n_{eff,z}$ for the anti-symmetric modes show negligible variation with gap sizes [see Fig. 3(c)], so that the repulsive optical forces for the anti-symmetric modes just increase slightly when the gap size shrinks [see Fig. 3(d)]. As a result, optical forces for the anti-symmetric mode are much weaker than that of symmetric modes, in sharp contrast to the case in dielectric slot waveguides, where the optical forces for the symmetric mode and the anti-symmetric mode are comparable in magnitude[4, 11, 28]. It is the strong interaction between the two waveguides that leads to the distinct mode coupling strengths and the distinguished optical forces obtained from the two eigenmodes. Furthermore, stronger optical forces are achieved in slot waveguides with a smaller cross section, for both the symmetric modes and the anti-symmetric modes, due to larger effective indices and thus stronger mode coupling strength.

In order to provide a comprehensive understanding of the relation between the gradient optical forces and the waveguide mode coupling, here we give an analytical expression to solve the optical forces in the metamaterials slot waveguides. Since the optical mode profiles (see Fig. 2) show a negligible dependence on the $x$ coordinate, the 3D coupled metamaterial waveguides can be approximately treated as 2D coupled slab waveguides. With the field components in the form of $\exp(ik_z z - i\omega t)$, the optical field of the slot waveguides can be expressed as

$$E_y = E_0 \begin{cases} n_{\text{eff},z} \cos\left(-k_y \dfrac{L_y}{2} + \varphi\right) \Phi(y), & 0 < y < \dfrac{g}{2} \\ \dfrac{n_{\text{eff},z}}{\varepsilon_y} \cos\left[k_y\left(y - \dfrac{L_y + g}{2}\right) + \varphi\right], & \dfrac{g}{2} < y < \dfrac{g}{2} + L_y \\ n_{\text{eff},z} \cos\left(k_y \dfrac{L_y}{2} + \varphi\right) \exp\left[-\gamma\left(y - \dfrac{g}{2} - L_y\right)\right], & y > \dfrac{g}{2} + L_y \end{cases} \quad (3)$$

where $\Phi(y) = \cosh(\gamma y)/\cosh(\gamma g/2)$ for the symmetric mode and $\Phi(y) = \sinh(\gamma y)/\sinh(\gamma g/2)$ for the anti-symmetric mode, $\varphi$ is the phase shift at the middle of each waveguide due to the mode coupling. The wave vector inside metamaterial $k_y$ and the field decay rate in air $\gamma$ are related to the propagation wave vector $k_z$ through the dispersion relations $\dfrac{k_z^2}{\varepsilon_y} + \dfrac{k_y^2}{\varepsilon_z} = k_0^2$ and $k_z^2 - \gamma^2 = k_0^2$ for hyperbolic metamaterial and air, respectively. Only the positive $y$ part of the whole expression is shown for clarity.

After substituting the optical fields into the Maxwell's stress tensor and taking into account that $n_{\text{eff},z} = k_z/k_0 \gg 1$, we obtain the following analytical expression for the optical forces in the slot waveguides of hyperbolic metamaterials:

$$f_{opt} \approx -\frac{1}{2cL_y} \frac{n_{\text{eff},z}}{\frac{1}{\varepsilon_y} + |\varepsilon_z|\sinh^2\left(\frac{\gamma g}{2}\right)}, \quad \text{symmetric mode}$$

$$f_{opt} \approx \frac{1}{2cL_y} \frac{n_{\text{eff},z}}{|\varepsilon_z|\cosh^2\left(\frac{\gamma g}{2}\right)}, \quad \text{anti-symmetric mode} \tag{4}$$

where $c$ is the speed of light in vacuum. Fig. 3 also plots the analytically derived effective indices $n_{\text{eff},z}$ and optical forces $f_{opt}$, which match the numerical FEM simulation results quite well. The critical dependence of the optical forces on the gap size $g$ can be explained using Eq. (4). The constructive (destructive) interference of exponentially decayed evanescent optical fields in the slot region gives rise to hyperbolic-sine-function (hyperbolic-cosine-function) dependence on gap sizes, resulting in strong (weak) transverse optical forces for the symmetric (anti-symmetric) modes as the gap size gets small. When the gap sizes becomes larger ($g > 20$ nm), the two waveguides can only couple weakly with each other through the tails of the evanescent optical fields, so that $f_{opt}$ becomes quite weak for both eigenmodes. According to Eq. (4), the waveguide height $L_y$, the mode indices $n_{\text{eff},z}$ and the permittivity $\varepsilon_y$ and $\varepsilon_z$ also affect the magnitude of optical forces.

The influence of permittivity tensors on the optical forces is shown in Fig. 4, by varying the silver filling ratio $f_m$ of hyperbolic metamaterials for slot waveguides of $L_y = 30$ nm with different gap sizes $g = 1$ nm and $g = 3$ nm. The effective indices $n_{\text{eff},z}$ at different filling ratios are shown in Fig. 4(a) and (c) for the symmetric modes and the anti-symmetric modes, respectively. It is noted that ultra-high refractive indices can be achieved for both low $f_m$ ($< 0.3$) and high $f_m$ ($> 0.75$) in the metamaterial slot waveguides. A high $f_m$ leads to a large $\varepsilon_y$, and a low $f_m$ gives a large ratio of $\sqrt{\varepsilon_y/|\varepsilon_z|}$ in hyperbolic metamaterials, both of which can result in high refractive indices along the propagation direction. However, the optical forces turn out to be large for metamaterial slot waveguides with low $f_m$, for both the symmetric and anti-symmetric modes, as shown in Fig. 4(b) and (d). This phenomenon can also be explained with Eq. (4), where optical forces are more sensitive to $|\varepsilon_z|$ (determined by the filling ratio $f_m$) other than the effective indices $n_{eff,z}$, due to the hyperbolic-function terms. Fig. 4 shows that the optical forces for the symmetric modes are always significantly higher than those for the anti-symmetric modes. Moreover, a smaller gap size ($g$ = 1 nm) corresponds to a stronger optical force compared with the case of a larger gap size ($g$ = 3nm). All these results are consistent with the previous calculation in Fig. 3.

With coupled mode theory (CMT)[12, 20, 29], the eigenmode $\psi$ supported in a coupled waveguide system can be expressed as the superposition of the two individual waveguide modes $\psi_1$ and $\psi_2$,

$$\psi = a_1\psi_1 + a_2\psi_2 \qquad (5)$$

where $a_1$ and $a_2$ ($|a_1|^2+|a_2|^2=1$) are the mode amplitudes of the first waveguide and the second waveguide, respectively. Due to the presence of mode coupling, the evolution of the eigenmode $\psi$ can be described by the following coupled mode equations,

$$n_1 a_1 + \kappa_\pm a_2 = n_{eff,z\pm} a_1$$
$$\kappa_\pm a_1 + n_2 a_2 = n_{eff,z\pm} a_2 \qquad (6)$$

where $n_{eff,z\pm}$ and $\kappa_\pm$ represent the effective indices and the coupling strengths for the symmetric waveguide mode (+) and the anti-symmetric waveguide mode (-), respectively. $n_1$ and $n_2$ are the effective indices of two individual waveguides. The coupling strength between the two waveguide modes can be derived as $\kappa_\pm = \sqrt{(n_{eff,z\pm} - n_1)(n_{eff,z\pm} - n_2)}$, which is closely related to the transverse optical forces[12]. Here we derive the formula between the optical forces $f_{opt}$ and the mode coupling strength $\kappa$ by exploiting another expression for the optical force calculation[11, 12],

$$f_{opt} = -\frac{1}{\omega v_g} \frac{\partial \omega}{\partial g}\bigg|_{k_z} \qquad (7)$$

which comes from the relation between the adiabatic variation of gap size and photonic energy of the coupled waveguide system. When the dispersion of group velocity $dv_g/d\omega$ is negligible, we have $\left.\frac{\partial \omega}{\partial g}\right|_{k_z} = -v_g \left.\frac{\partial k_z}{\partial g}\right|_{\omega}$ so that the Eq. (7) can be rewritten as,

$$f_{opt} = \frac{1}{\omega v_g} v_g \left.\frac{\partial k_z}{\partial g}\right|_{\omega} = \frac{1}{\omega}\left.\frac{\partial(n_{eff,z} k_0)}{\partial g}\right|_{\omega} = \frac{1}{c}\left.\frac{\partial n_{eff,z}}{\partial g}\right|_{\omega} \qquad (8)$$

It should be noted that the partial derivative in Eq. (8) is performed at a fixed frequency, while the partial derivative in Eq. (7) is performed at a fixed propagation wave vector. After the substitution of $n_{eff,z}$ in Eq. (8) with the expression for the mode coupling strength $\kappa$, the relation between the optical force and the coupling strength is obtained,

$$f_\pm = \frac{1}{c}\frac{\partial n_{eff,z\pm}}{\partial g} = \frac{1}{c}\frac{1}{\pm\sqrt{\cot^2\theta+1}}\frac{\partial \kappa_\pm}{\partial g} = \pm\frac{\sin\theta}{c}\frac{\partial \kappa_\pm}{\partial g} \qquad (9)$$

where $\theta = \tan^{-1}\left(\frac{2\kappa_\pm}{|n_1-n_2|}\right)$ and $f_\pm$ corresponds to the optical forces for the symmetric modes (+) and the anti-symmetric modes (-), respectively. In the coupled mode theory, $\sin^2(\theta)$ corresponds to the maximum power transfer efficiency from one waveguide to the other in a coupled waveguide system[20]. For a coupled system with two identical waveguides, $\sin(\theta) = 1$, and Eq. (9) is reduced to $f_\pm = \pm\frac{1}{c}\frac{\partial \kappa_\pm}{\partial g}$. This formula reveals that the transverse optical forces are proportional to the variation rate of the mode coupling strength as the two waveguides approach each other adiabatically. To check the validity of this formula, the calculated optical forces using Eq. (9) are displayed in Fig. 5. It is shown that the CMT formula can give optical forces exactly the same as that rigorously calculated from the integration of Maxwell's stress tensor for both the symmetric modes and the anti-symmetric modes.

Finally, a realistic hyperbolic metamaterial slot waveguide is constructed using alternative silver and germanium layers with a period of 10 nm and a silver filling ratio of $f_m = 0.4$. The comparison between the multilayer structures and the ideal effective medium is shown in Fig. 6. With the height of $L_y = 80$ nm, the slot waveguides of multilayer structures can reproduce the results of the effective indices $n_{eff,z}$ and the optical forces $f_{opt}$ calculated based on the waveguides of ideal effective media, for both the symmetric modes and the anti-symmetric modes. While with $L_y$ = 30 nm, both $n_{eff,z}$ and $f_{opt}$ become smaller than the results of the effective media calculation, which is due to that the large wave vector along the $y$ direction becomes close to the Brillouin zone of periodic multilayer structures, so that the waveguide mode profiles begin to deviate from those predicted from the effective medium theory.

In conclusion, we have demonstrated giant transverse optical forces up to 8 nNμm$^{-1}$mW$^{-1}$ in nanoscale slot waveguides of hyperbolic metamaterials, due to the strong optical field confinement and extreme optical energy compression within the air slot region. The influences of the waveguide geometries and the metal filling ratios of the hyperbolic metamaterials on both the symmetric modes and the anti-symmetric modes are studied numerically using the Maxwell stress tensor integration method, together with an analytical approach using a 2D approximation of the 3D coupled waveguides. Furthermore, the relation between the transverse optical forces and the waveguide mode coupling strength is derived from the coupled mode theory, revealing the mechanism of optical force enhancement in slot waveguide system. Finally, it is shown that the predicted giant optical force is achievable in realistic metal-dielectric multilayer structures. The strongly enhanced optical forces in slot waveguides of hyperbolic metamaterial will open a new realm in many exciting nanoscale optomechanical applications.

**Acknowledgements**

This work was partially supported by the Department of Mechanical and Aerospace Engineering, the Materials Research Center, the Intelligent Systems Center, and the Energy Research and Development Center at Missouri S&T, the University of Missouri Research Board, the Ralph E. Powe Junior Faculty Enhancement Award, and the National Natural Science Foundation of China (61178062 and 60990322).

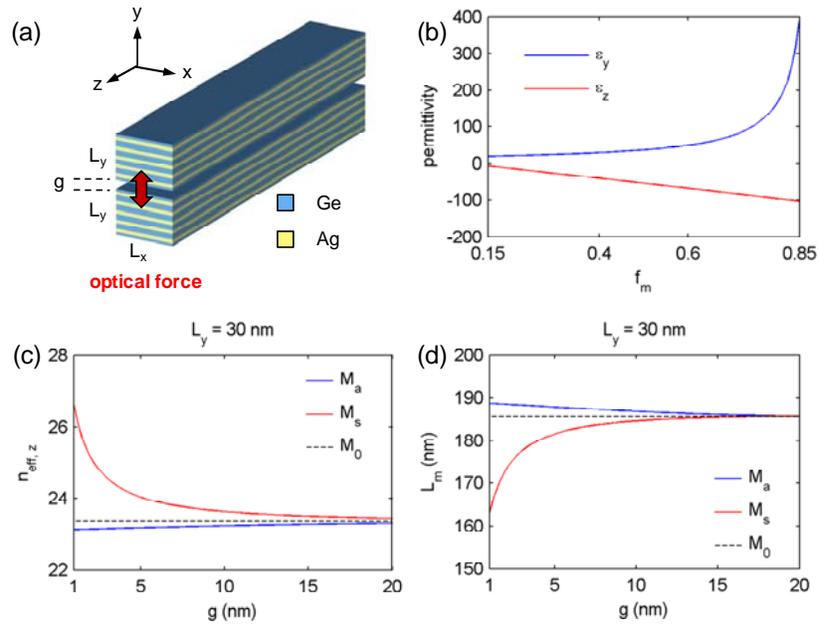

**Figure 1.** (a) The schematic of nanoscale slot waveguides of hyperbolic metamaterials. Giant transverse optical forces are expected to occur due to the strong coupling between the two waveguides. (b) The dependence of the effective permittivity tensor on the silver filling ratio $f_m$. (c) The effective refractive indices along the propagation direction $n_{eff,z}$ and (d) the propagation length $L_m$ for slot waveguide modes as functions of gap sizes $g$. In (c) and (d), $M_a$, $M_s$ and $M_0$ represent the anti-symmetric mode, the symmetric mode and the unperturbed mode of an individual waveguide, respectively.

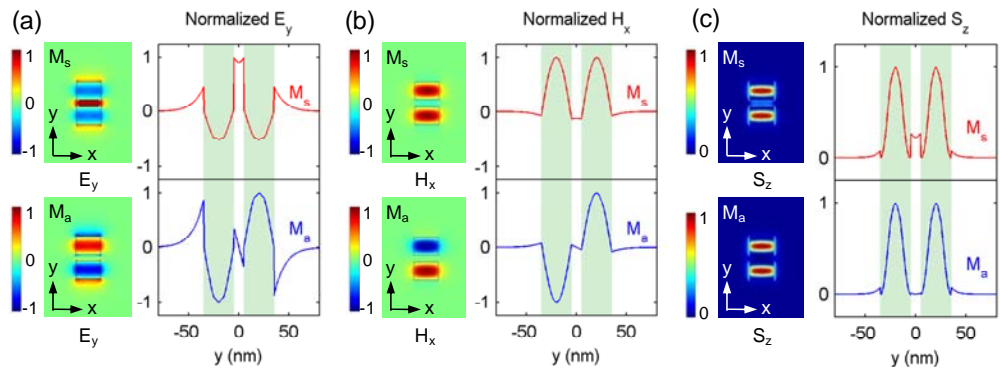

**Figure 2.** The 2D optical mode profiles of (a) $E_y$, (b) $H_x$ and (c) $S_z$ for metamaterial slot waveguide with $g = 10$ nm. The crossing line plots along $x = 0$ are also shown for clarity. In each panel, the optical mode profiles for the symmetric mode $M_s$ are plotted at the top and the profiles for the anti-symmetric mode $M_a$ are plotted at the bottom.

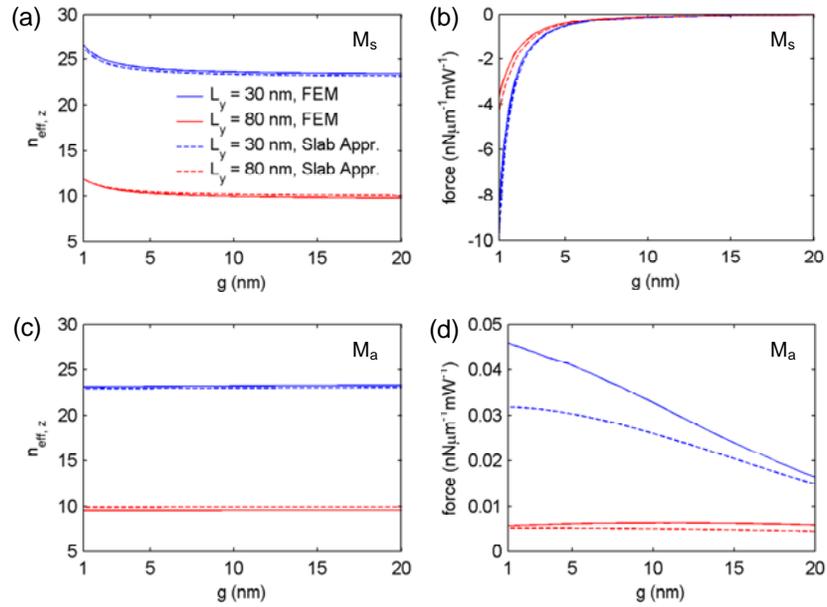

**Figure 3.** The calculated effective refractive indices $n_{\text{eff},z}$ and optical forces $f_{\text{opt}}$ in metamaterial slot waveguides with different gap sizes $g$, (a) and (b) for the symmetric modes, and (c) and (d) for the anti-symmetric modes. Results from FEM simulations (FEM) are plotted in solid lines, and results from 2D coupled slab waveguide approximation (Slab Appr.) are plotted in the dashed lines. Two waveguide heights $L_y = 30$ nm and $L_y = 80$ nm are shown for comparison.

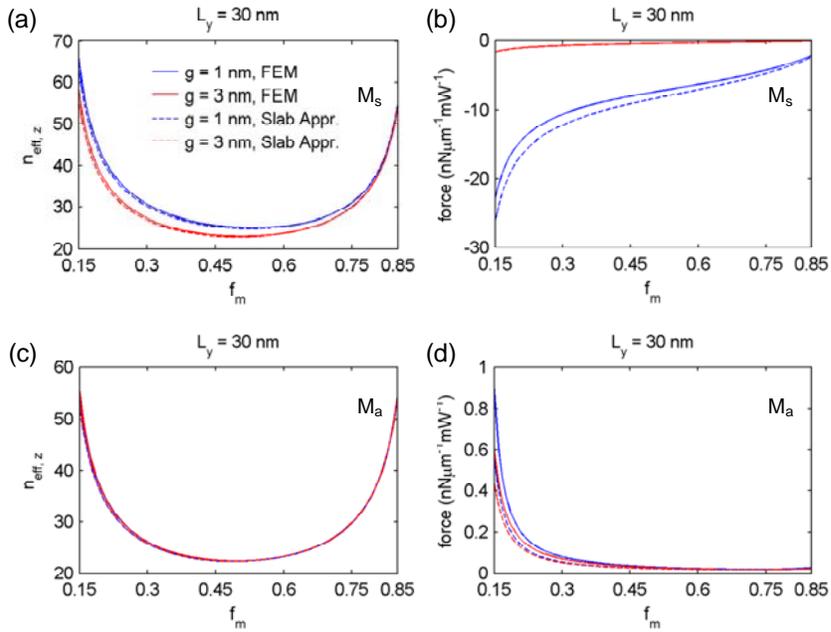

**Figure 4.** The calculated effective refractive indices $n_{\text{eff},z}$ and optical forces $f_{\text{opt}}$ for the metamaterial slot waveguides with different silver filling ratios $f_m$, (a) and (b) for the symmetric modes, and (c) and (d) for the anti-symmetric modes. Results from FEM simulation (FEM) are plotted in solid lines, and results from 2D coupled slab waveguide approximation are plotted in the dashed lines. Two waveguide gap sizes, $g = 1$ nm and $g = 3$ nm, are shown for comparison.

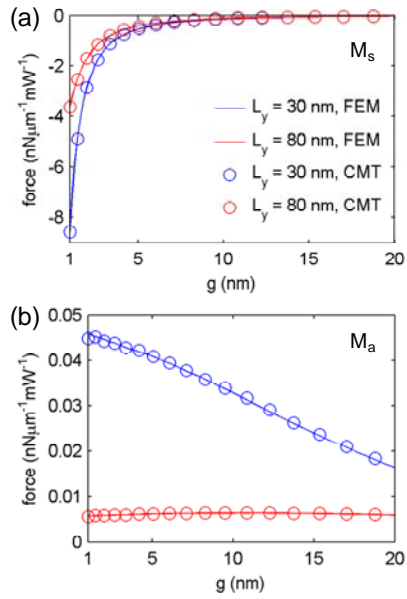

**Figure 5.** The optical forces $f_{opt}$ of (a) the symmetric modes and (b) the anti-symmetric modes calculated based on coupled mode theory (CMT) analysis (in circles) for metamaterial slot waveguides with different gap sizes, which perfectly match the results from rigorous FEM simulations (in solid lines).

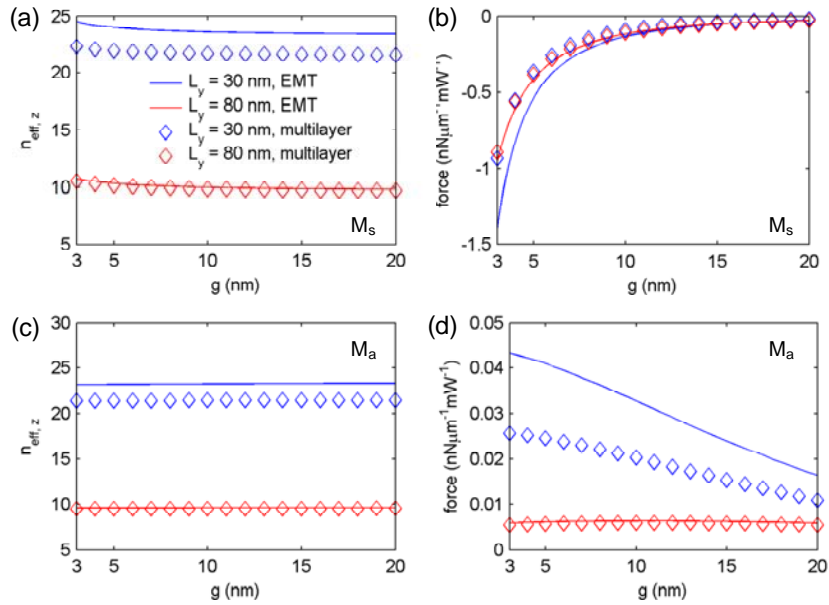

**Figure 6.** The calculated effective refractive indices $n_{\text{eff},z}$ and optical forces $f_{\text{opt}}$ on the realistic metal-dielectric multilayer structures (in diamonds) with the layer pitch of 10 nm and the filling ratio of 0.4 at different gap sizes, (a) and (b) for the symmetric modes and (c) and (d) for the anti-symmetric modes. Results from ideal effective medium theory (EMT) are plotted in solid lines for comparison.